\documentclass[11pt,a4paper]{article}
\pdfoutput=1
\usepackage[hyperref]{emnlp-ijcnlp-2019}
\usepackage{times}
\usepackage{latexsym}
\usepackage{listings}
\usepackage{graphicx}
\usepackage{subfig}
\usepackage{tabularx}
\usepackage{bbm}
\usepackage{pifont}
\usepackage{svg}

\newcommand{\greencheckmark}[0]{\ding{52}}
\usepackage{xcolor}
\usepackage{xspace}

\graphicspath{ {./images/} }
\usepackage{xcolor}

\definecolor{codegreen}{rgb}{0,0.6,0}
\definecolor{codegray}{rgb}{0.5,0.5,0.5}
\definecolor{codepurple}{rgb}{0.58,0,0.82}

\lstdefinestyle{mystyle}{
    commentstyle=\color{codegreen},
    keywordstyle=\color{magenta},
    numberstyle=\tiny\color{codegray},
    stringstyle=\color{codepurple},
    basicstyle=\ttfamily\footnotesize,
    breakatwhitespace=false,         
    breaklines=true,                 
    captionpos=b,                    
    keepspaces=true,                 
    numbers=left,                    
    showspaces=false,                
    showstringspaces=false,
    showtabs=false,                  
    tabsize=2, 
    frame=single
}

\usepackage{url}

\aclfinalcopy 

\newcommand\sciwing{SciWING\xspace}
\newcommand{\sciwinglogo}{\raisebox{-.15\height}{\includegraphics[height=4\fontcharht\font`\B]{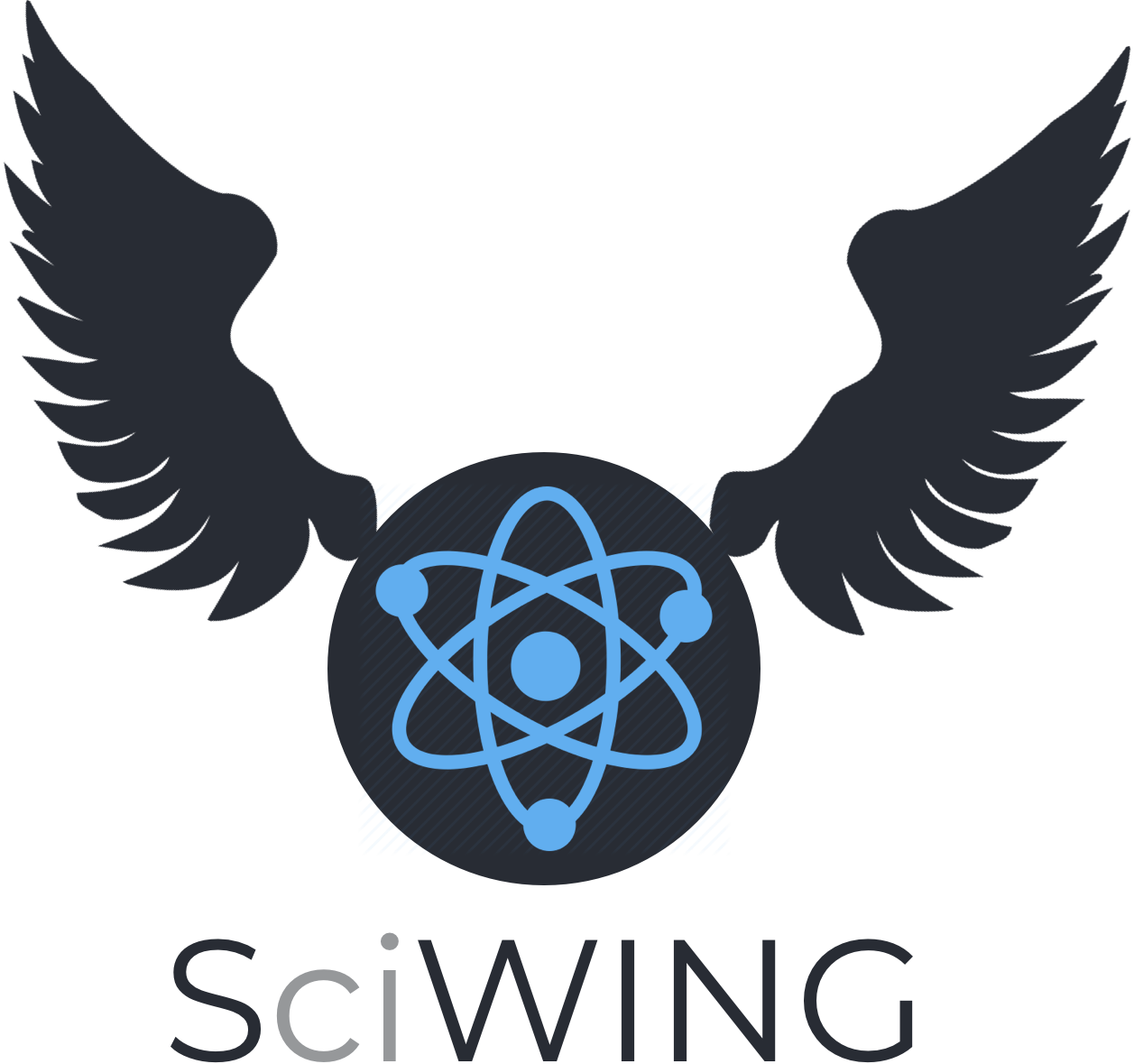}}}

\title{\sciwinglogo{} {\normalsize SciWING} -- A Software Toolkit for Scientific Document Processing}

\author{Abhinav Ramesh Kashyap \\
  National University of Singapore \\ 
  {\tt abhinav@comp.nus.edu.sg} \\\And
  Min-Yen Kan \\
  National University of Singapore \\
  {\tt knmnyn@comp.nus.edu.sg} \\}

\date{}

\begin{document}

\maketitle



\begin{abstract}
We introduce \sciwing{}, an open-source software toolkit which provides access to state-of-the-art pre-trained models for scientific document processing (SDP) tasks, such as citation string parsing, logical structure recovery and citation intent classification. Compared to other toolkits, \sciwing{} follows a full neural pipeline and provides 
a Python interface for SDP. When needed, \sciwing{} provides fine-grained control for rapid experimentation with different models by swapping and stacking different modules.  Transfer learning from general and scientific documents specific pre-trained transformers (i.e., BERT, SciBERT, etc.) can be performed. \sciwing{} incorporates ready-to-use web and terminal-based applications and demonstrations to aid adoption and development. 
The toolkit is available from \url{http://sciwing.io} and the demos are available at \url{http://rebrand.ly/sciwing-demo}\footnote{Watch our demo video at \url{https://rebrand.ly/sciwing-video}}.
\end{abstract}

\lstset{linewidth=7.5cm,xleftmargin=2em}

\section{Introduction}
Automated scientific document processing (SDP) deploys natural language processing (NLP) on scholarly articles.  As scholarly articles are long-form, complex documents with conventional structure and cross-reference to external resources, they require specialized treatment and have specialized tasks.  Representative SDP tasks include parsing embedded reference strings \cite{prasadanimesh,thai2020using}; identifying the importance, sentiment and provenance for citations \cite{Cohan2019Structural,Su2019NeuralML}; identifying logical sections and markup \cite{luong2012logical}; parsing of equations, figures and tables \cite{Clark2016PDFFigures2M}; and article summarization \cite{qazvinian-radev-2008-scientific,qazvininaradev2013,cohan-goharian-2015-scientific,cohan-etal-2018-discourse,cohanijdl}.  SDP tasks, in turn, help downstream systems and assist scholars in finding relevant documents and manage their knowledge discovery and utilization workflows. 
Next-generation introspective digital libraries such as Semantic~Scholar~\cite{ammar-etal-2018-construction}, Google Scholar and Microsoft Academic have begun to incorporate such services. 

\begin{figure}[t!]
    \centering
    \includegraphics[width=0.5\textwidth]{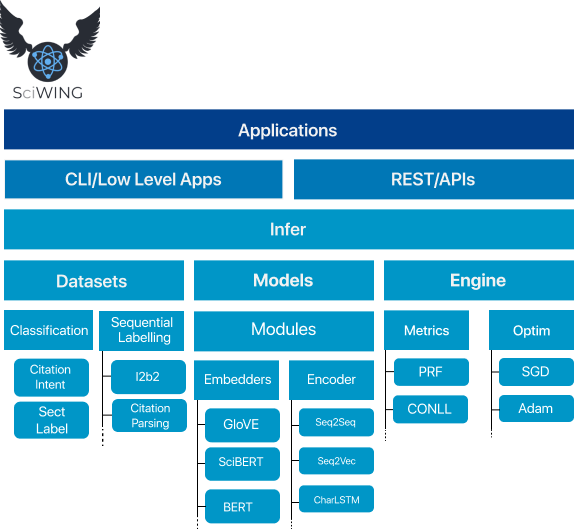}
    \caption{\sciwing{} Components: Text classification 
     and Sequence labelling {\bf Datasets}, {\bf Models} composed from low-level {\bf Modules}, and \textbf {Engine} to train and record experiment parameters. {\bf Infer} middleware does the inference and high-level functionality (e.g., developing APIs; low-level and web applications).}
    \label{fig:sciwing_arch}
\end{figure}

\begin{table*}[t!]
    \centering
    \begin{tabular}{c c c c c c}
    \textbf{Framework} & \textbf{Pre-trained models} & \textbf{SOTA} &\textbf{ Neural first} & \textbf{Extensible} & \textbf{Language/Framework} \\
    \hline
    SciSpaCy & \greencheckmark &  & & & Py \\ 
    AllenNLP & \greencheckmark & & \greencheckmark & \greencheckmark & Py(Torch) \\ 
    \textsc{flair} & \greencheckmark & & \greencheckmark & \greencheckmark & Py(Torch) \\
    Grobid &  \greencheckmark & & & & Java \\ 
    \hline 
    \sciwing{} & \greencheckmark & \greencheckmark & \greencheckmark & \greencheckmark & Py(Torch) \\ 
    \hline
    \end{tabular}
    \caption{Comparison of \sciwing{} with popular frameworks. \textbf{Pre-trained models}: availability of pretrained models. \textbf{SOTA}: state-of-the-art or comparably performing models for SDP. \textbf{Neural-Networks first}: supports end-to-end neural network development and training. \textbf{Extensible}: easily incorporates new datasets and architectures.}
    \label{tab:framework-comparison}
\end{table*}

While NLP, in general, has seen tremendous progress with the introduction of neural network architectures and general toolkits and datasets to leverage them, their deployment for SDP is still limited.  Over the past few years, many open-source software packages have accelerated the development of state-of-the-art (SOTA) NLP models. However, these frameworks have a few limitations concerning SDP. First, most are general purpose frameworks aimed at producing SOTA models for natural language understanding tasks or only for specific domains such as biomedicine. Second, they do not provide deployable, SOTA models for SDP. Most provide limited or no means for researchers to train models on their datasets, or experiment with model architectures.



A key barrier to entry is accessibility: a non-trivial level of expertise in NLP and machine learning is a prerequisite.  Practitioners who wish to deploy SDP on their field's literature may lack knowledge and motivation to learn it for the sake of deployment.  Thus, there is a clear need for a toolkit that unifies different efforts and  provides access to pre-trained, SOTA models for SDP, while also allowing researchers to experiment with models rapidly to create deployable applications.

We introduce \sciwing{} to address this need with respect to other frameworks (cf. Table~\ref{tab:framework-comparison}). Built on top of PyTorch, 
it provides access to neural network models 
pre-trained
for a growing number of SDP tasks which practitioners can easily deploy on their documents.
For researchers, these models serve as baselines and \sciwing{} encourage them to try architectural variations in a modular manner by swapping or composing 
modules 
in a declarative manner in a configuration file, 
without having to program or deal with complexities of generic tools like AllenNLP \cite{Gardner2017AllenNLP}. 

\sciwing{} is MIT licensed and comes with downloadabled pre-trained models and freely-available datasets. The package runs on Python 3.7 and can be easily installed from Python Packaging Index (PyPI) using \texttt{pip install sciwing}. For researchers aiming to further develop \sciwing{}, we provide installation tools that set up the system, alongside 
documentation. 
\section{System Overview}

Our view is that SDP-specific considerations are best embodied as an abstraction layer over existing NLP frameworks.  \sciwing{} incorporates AllenNLP, the generic NLP pipeline \cite{Gardner2017AllenNLP}, developing models on top of it when necessary, while using the transformers package \cite{Wolf2019HuggingFacesTS} to enable transfer learning via its pre-trained general-purpose representations such as BERT \cite{devlin-etal-2019-bert} and SDP-specific ones like SciBERT \cite{beltagy-etal-2019-scibert}. 
Fig.~\ref{fig:sciwing_arch} shows 
\sciwing{}'s Dataset, Model and Engine components facilitating flexible re-configuration.  We now describe these components.



    \vspace{2mm}
    \textbf{Datasets}: There are many challenges for the researcher--practitioner to experiment with different SDP tasks.  First, 
    researchers must handle various data formats and datasets: for reference string parsing, the CoNLL format is most common; for text classification, CSV is most common. \sciwing{} enables reading of dataset files in different formats and also facilitates the download of open datasets using command-line interfaces. For example, \texttt{sciwing download data --task scienceie} downloads the official, openly-available dataset for the ScienceIE task.  
    
    Current methods for pre-processing are cumbersome and error-prone.  Processing can become complex when different models require different tokenisation and numericalisation methods.  \sciwing{} unifies these various input formats through a pipeline of pre-processing, tokenisation and numericalisation, via \texttt{Tokenisers} and \texttt{Numericalisers}. \sciwing{} also handles batching and padding of examples.
    \vspace{1mm}
    
    \textbf{Models}: Paired Embedder--Encoder subcomponents combine to form a neural network model, themselves PyTorch classes. 
    
    {\it Embedders}: Modern NLP models represent natural language tokens as continuous vectors -- embeddings.  \sciwing{} abstracts this concept via \texttt{Embedders}.  Generic (non-SDP specific) embeddings such as GlovE \cite{pennington-etal-2014-glove} are natively provided.  Tokens in scientific documents can benefit from special attention, as most are missing from pre-trained embeddings. \sciwing{} includes task-specific trained embeddings for reference strings \cite{prasadanimesh}. \sciwing{} also supports contextual word embeddings: ELMo~\cite{peters-etal-2018-deep}, BERT~\cite{devlin-etal-2019-bert}, SciBERT~\cite{beltagy-etal-2019-scibert}, etc. SOTA embedding models are built by concatenating multiple representations, via \sciwing{}'s \texttt{ConcatEmbedders} module. As an example, word and character embeddings are combined in NER models~\cite{lample-etal-2016-neural}, and multiple contextual word embeddings are combined in various clinical and BioNLP tasks \cite{zhai-etal-2019-improving}.
    
    {\it Neural Network Encoders}: \sciwing{} ports commonly-used neural network components that can be composed to form neural architectures for different tasks. For example in text classification, encoding input sentence as a vector using an LSTM is a common task 
    (\sciwing{}'s 
    \texttt{seq2vecencoder}). Another common operation is obtaining a sequence of hidden states for a set of tokens, often used in sequence labelling tasks and 
    \sciwing{}'s 
    \texttt{Lstm2seq} achieves this. Further, it also includes attention based modules.
    
    \sciwing{} builds in generic linear classification and sequence labelling with CRF heads that can be attached to the encoders to build production models. It provides pretrained SOTA models for particular SDP tasks that work out-of-the-box 
    or further fine-tuned.  
    
    \vspace{2mm}
    
    \textbf{Engine}: 
    \sciwing{} handles all the boilerplate code to train the model, monitor the loss and metrics, check-pointing parameters at different stages of training, validation and testing. It helps researchers adopt best practices, such as clipping gradient-norms, as well as saving and deploying best performing models. 
    Users can customize the following: \\

    {\it Optimisers}: \sciwing{} supports all the optimisers supported by PyTorch, and various learning rate schedulers that dynamically manage learning rates based on validation performance.
    
    {\it Experiment Logging}:
    \sciwing{} adopts current best practices in leveraging logging tools to monitor and manage experiments. \sciwing{} writes logs for every experiment run and facilitates cloud-based experiment logging and corresponding charting of relevant metrics via the third-party API service of {\it Weights and Biases}\footnote{\url{www.wandb.com}.}, with the integration of alternative logging services on the way.
    
    {\it Metrics}: Different SDP tasks require their respective metrics. \sciwing{} abstracts a separate \texttt{Metrics} module to select appropriate metrics for each task. \sciwing{} includes \texttt{PrecisionRecallFMeasure} suitable for text classification tasks,  \texttt{TokenClassificationAccuracy}, and the official CONLL2003 shared task evaluation metric
    suitable for sequence labelling. \\
    
\noindent With these components given, \sciwing{}'s \textbf{Inference} middleware provides clear abstractions to perform inference once models are trained. The layer runs predictions on the test dataset, user inputs and files.  Such abstractions also act as an interface for the development of upstream REST APIs and command-line applications. 
    
\subsection{Configuration using TOML} 

\sciwing{}'s flexible architecture is encapsulated by its use of declarative TOML configuration files.  TOML was chosen as it is a widely-used, unambiguous, and human-readable configuration file format.
This enables users to declare dataset, model architectures and experiment hyper-parameters in a single place. \sciwing{} parses the TOML file and creates appropriate instances of the dataset, model and engine to run experiments.

A simple configuration file for reference string parsing  along with its equivalent model declaration in Python is shown in the listings below. The class declaration 
and configuration file have a one-to-one correspondence.  As deep learning models are made of multiple modules, \sciwing{} 
automatically instantiates these submodules as needed. \sciwing{} constructs a Directed Acyclic Graph (DAG) from the model definition to achieve this. The DAG's topological ordering instantiates the different submodules to form the final model.\\

\begin{lstlisting}[language=XML, basicstyle=\small]
[model]
    class="SimpleClassifier"
    encoding_dimension=300
    num_classes=23
    classification_layer_bias=true
    [model.encoder]
        emb_dim=300
        class="BOW_Encoder"
        dropout_value=0.5
        aggregation_type="sum"
        [[model.encoder.embedder]]
        class="VanillaEmbedder"
        embed="word_vocab"
        freeze=False
\end{lstlisting}

\begin{lstlisting}[language=Python, basicstyle=\small]
class SimpleClassifier(nn.Module):
    def __init__(
    self, 
    encoder: nn.Module,
    encoding_dim: int,
    num_classes: int,
    classification_layer_bias: bool)
\end{lstlisting}


\subsection{Command Line Interface}
\label{sec:command-line-interfaces}
Qualitatively analyzing the results of the model by drilling down to certain training and development instances can be telling and help to diagnose performance issues. \sciwing{} facilitates this by providing an interactive inspection of the model  through a command-line interface (CLI). Consider the task of reference string parsing: the confusion matrix for the different classes can be displayed through the provided CLI utility, which also allows finer-grained introspection of (Precision, Recall, F-measure) metrics and the viewing of error instances where one class is confused for another. For example, \texttt{sciwing interact neural-parscit} provides introspection utilities for the pre-trained reference string parsing model. Such introspection utilities are also available for other pre-trained models.

\sciwing{} provides commands to run experiments from the configuration file, aiding replication. For example, experiments declared in a file named \texttt{experiment.toml}, can be run with the command \texttt{sciwing run experiment.toml}. \sciwing{} then saves the best model. \texttt{sciwing test experiment.toml} invokes inference which deploys the best model against the test dataset and displays the resultant metrics.

\subsection{End User Interfaces}
\sciwing's API service enables the development of various graphical user interfaces. \sciwing{} uses its \textit{Infer} layer and exposes APIs for various tasks including reference string parsing, citation intent classification, extracting abstracts and logical sections of articles, identifying entities in clinical notes, using \texttt{fastapi}\footnote{https://fastapi.tiangolo.com/}.
The API enables the following application families downstream:\\

\begin{figure*}
    \centering
    \includegraphics[width=\textwidth]{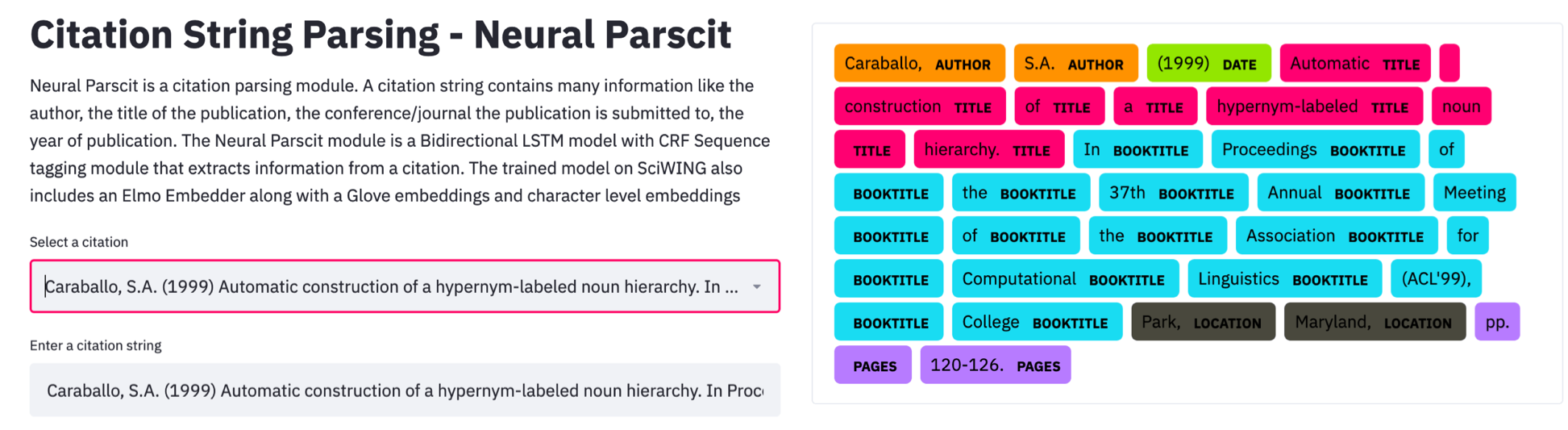}
    \caption[]{Sample \sciwing{}'s demonstration (\url{https://rebrand.ly/sciwing-demo}) for reference string parsing model, where input (l) is then classified into 13 output classes (r). We utilize the displaCy visualization toolkit (www.spacy.io) and  streamlit (www.streamlit.io).  }
    \label{fig:parscit-demo}
\end{figure*}

     \textbf{$\bullet$ Web Demonstrations}: To provide quick access to predictions from state-of-the-art models, fulfilling one key aim of \sciwing{}, we have developed an interactive demo using streamlit\footnote{http://www.streamlit.io}:  \url{http://rebrand.ly/sciwing-demo}. Pre-specified data or user data can be  processed using the distributed models (Figure~\ref{fig:parscit-demo}). Both API services and demos can also be run by installing \sciwing{} locally.
    
    \textbf{$\bullet$ Programmatic Interfaces} in \sciwing{} provisions advanced use. Users can make predictions for documents stored as .pdf or text files. For example, to parse a text file's citations, \sciwing provides a \texttt{NeuralParscit} class that has methods to parse all the strings in a file, storing them in a new file. Such a programmatic interface helps the practitioner make predictions easily.

\begin{table}[t]
    \centering
    \small
    \begin{tabular}{l | c | c}
        \hline
         \textbf{Task} & \textbf{SciWING} & \textbf{Best} \\
         \hline 
         Reference String Parsing & \textbf{88.44} & --- \\
         ScienceIE & \textbf{49.9} & 48.01 \\ 
         Logical Structure Recovery & \textbf{73.2} & --- \\ 
         Citation Intent Classification & 82.16  & 82.6 \\ 
         I2B2 NER & 85.83 & 86.23 \\
         \hline 
    \end{tabular}
    \caption{\sciwing{}'s SDP task performance, compared against other comparable models. 
    Bolded values indicate state-of-the-art or comparable performance (without attention).
    Scores are macro $F_1$.}
    \label{tab:sdp_performance}
\end{table}

\section{Tasks}
\sciwing{} prepackages models for various SDP tasks.  The examples demonstrate how to use the framework effectively. These models have performance close or comparable to state-of-the-art models (Table~\ref{tab:sdp_performance}).  They are production-ready, but also can be used as baselines for further research.\\

    \textbf{$\bullet$ Reference String Parsing} assigns one of 13 classes to tokens of a reference string that correspond with a in-document citation: {\it author}, {\it journal} and {\it year} of publication, among them.  Neural sequence labelling models, combining a bidirectional LSTM with CRF currently yield top results \cite{prasadanimesh}. The model included in \sciwing{} implements the same model architecture, but adds ELMo embeddings.  Unfortunately, due to the high training expense of their 10-fold cross validation, we are not able to obtain directly comparable results to their model's performance.

    \textbf{$\bullet$ ScienceIE} identifies typed keyphrases, originally from chemical documents: {\it Task} keyphrases that denote the end task or goal, {\it Material} keyphrases indicate any chemical, and {\it Dataset} that is being used by the scientific work and the process includes any scientific model or algorithm.
    The state-of-the-art system from 2017 includes a word and character embeddings and a bidirectional LSTM with CRF and uses language model (LM) embeddings \cite{ammar-etal-2017-ai2}. \sciwing{} includes a reference implementation without using LM embeddings and the results are comparable. We use the same dataset used by \cite{luong2012logical} for training the neural networks.

    \textbf{$\bullet$ Logical Structure Recovery} identifies the logical sections of a document: introduction, related work, methodology, and experiments.  This drives the relevant, targeted text to downstream tasks such as summarization, citation intent classification, among others. Currently, there are no neural network methods for this task, so \sciwing{}'s models can serve as strong baselines.

    \textbf{$\bullet$ Citation Intent Classification} identifies the purpose of a citation. Some citations refer to another work for {\it background} knowledge, a few to a related work's {\it results} and others to {\it compare and contrast} their methods or results. Such citation intents get used in Semantic Scholar\footnote{www.semanticscholar.org}. We train a
    bi-LSTM with ELMo on the Scicite dataset and achieve an F-score of 82.16. \cite{Cohan2019Structural} use Bi-LSTM with attention   At the time of writing, \sciwing \ does not include attention models, and as such results are not strictly comparable. We plan to include attention-based models in future iterations of the package.
    
    \textbf{$\bullet$ I2B2 Named Entity Recognition} identifies three kinds of entities from clinical notes; problems (e.g., a disease), treatments (e.g., a drug) and tests (e.g., diagnostic procedures). We use a similar model of Bi-LSTM CRF with ELMo embeddings, achieving 85.83 $F_1$. 
 


\section{Use Cases}

\sciwing{} caters to both use cases of practitioners looking to deploy pre-trained models as well as researchers looking to refine model architectures and perform fine-tuning domain adaptation on top of state-of-the-art contextual word embedding models.  We now examine both use cases.

\subsection{Using Pre-trained Reference String Parser}
\sciwing{} provisions out-of-the-box access to pretrained models for direct deployment. Citation string parsing can be deployed with just a few lines of code as shown below.

\begin{lstlisting}[language=Python, basicstyle=\small]
from sciwing.models.neural_parscit import NeuralParscit

# instantiate the best model for reference string parsing.
neural_parscit = NeuralParscit()

# predict for some reference
neural_parscit.predict_for_text("reference")

# predict for a file containing one reference per line 
neural_parscit.predict_for_file(/path/to/file)
\end{lstlisting}


\subsection{Building a Reference String Parser from Scratch}

State-of-the-art models can be built by stacking up multiple components. We illustrate how to construct such \sciwing{} models, building up to such SOTA model by simple modifications. Such step-by-step model creation also facilitates ablation studies, a common part of empirical studies.
\vspace{1mm}

\textbf{1. Bi-LSTM tagger}:
Our base model is a bi-LSTM with a  GLoVE embedder.  Every input token is classified into one of 13 different classes.
\begin{lstlisting}[language=Python, basicstyle=\small]
# initialize a word embedder 
word_embedder = WordEmbedder(
    embedding_type = "glove_6B_100")

# initialize a LSTM2Seq encoder
lstm2seqencoder = LSTM2SeqEncoder(
    embedder = word_embedder, 
    hidden_dim = 100,
    bidirectional = True)

# initialize a tagger without CRF
model = SimpleTagger(
    rnn2seqencoder = lstm2seqencoder, encoding_dim = 200)
\end{lstlisting}

\textbf{2. Bi-LSTM Tagger with CRF}:
We then make a single modification to the above code, swapping the simple tagger with one that uses a CRF.
\begin{lstlisting}[language=Python, basicstyle=\small]
...

# an RNN tagger with CRF on top
model = RnnSeqCrfTagger(
    rnn2seqencoder = lstm2seqencoder, encoding_dim = 200
)
\end{lstlisting}

\textbf{3. Bi-LSTM tagger with character and ELMo Embeddings}:
We modify the code to include a bidirectional LSTM character embedder.  We use the ConcatEmbedders module to create the final word embeddings (Line~16), which concatenates the character embeddings with those from the previous word embedding and a pretrained ELMo contextual word embedding. 
This final model is the provisioned model for the reference string parsing task provided in \sciwing{}.

\begin{lstlisting}[language=Python, basicstyle=\small]
...
word_embedder = WordEmbedder(
   embedding_type = "glove_6B_100"
)

# LSTM character embedder
char_embedder = CharEmbedder(
   char_embedding_dimension = 10,
   hidden_dimension = 25,
)

# ELMo embedder
elmo_embedder = ElmoEmbedder()

# Concatenate the embeddings
embedder = ConcatEmbedders([word_embedder, char_embedder, elmo_embedder])
\end{lstlisting}

Here we have described one use case, but models for other tasks are built as easily. In a similar way, we have built a similar architecture for the clinical notes parsing task --- the ScienceIE task I2B2 NER task.
\section{Related Work}
\label{sec:related-work}
Grobid~\cite{GROBID} is the closest to a general workbench for scientific document processing. 
Similarly to \sciwing{}, Grobid also performs document structure classification, reference string parsing, among other tasks. But Grobid is architected in the traditional, manual feature engineering approach, leading to performance losses for many SDP tasks, and difficulties in retrofitting neural models into its framework.   

SciSpaCy~\cite{Neumann2019ScispaCyFA} focuses on biomedical related tasks such as POS-tagging, syntactic parsing and biomedical span extraction.  However, SciSpaCy primarily caters for practitioners; it does not easily allow for the development and testing of new models and architectures.  

Task- and domain-agnostic frameworks also exist.  NCRF++ \cite{yang2018ncrf} is a tool for performing sequence tagging using Neural Networks and Conditional Random Fields and \textsc{flair} \cite{akbik2018coling} is a framework for general-purpose NLP and mainly provide access to different embeddings and ways to combine them.


\section{Conclusion and Future Work}

We introduce \sciwing{}, an open-source scholarly document processing (SDP) toolkit, targeted at practitioners and researchers interested in rapid experimentation.   It provisions pre-trained models for key SDP tasks 
that achieve state-of-the-art performance and aids practitioners to deploy models directly on their community's literature.

\sciwing{}'s modular design also greatly facilitates SDP researchers in model architecture development, speeding train/test cycles for architecture search, and supporting transfer learning for use cases with limited annotated data. \sciwing{} allows declaration of models, datasets and experiment parameters in a single configuration file. 

\sciwing{} is actively being developed. We consider the following improvements in our roadmap:

\textbf{$\bullet$} \sciwing{} has yet to incorporate natural language generation related models. We would like to consider sequence to sequence neural models which have proven useful for scientific document summarization tasks, among others.

\textbf{$\bullet$} Scientific document processing involves minimal training data and has found benefits in incorporating document structure, both of which are tackled using multi-task learning. Multi-task learning is thus a future milestone in \sciwing{}.

\textbf{$\bullet$} We would like \sciwing{} to foster collaboration among the SDP community and encourage assistance with these goals through contributions to our Github repository in the form of models, datasets and improvements to the framework.



\bibliography{emnlp-ijcnlp-2019}
\bibliographystyle{acl_natbib}

\appendix
\end{document}